\documentclass[12pt]{article}

\usepackage[
  left=0.8in,
  right=0.8in,
  top=0.6in,
  bottom=0.8in
]{geometry}

\usepackage[utf8]{inputenc}
\usepackage[T1]{fontenc}
\usepackage[english]{babel}
\usepackage{nameref}
\usepackage{todonotes}
\usepackage{graphicx}
\usepackage{subcaption}
\usepackage[font=small,labelfont=bf]{caption}
\usepackage{ragged2e}
\usepackage{float}
\usepackage{adjustbox}

\usepackage{amsmath}
\usepackage{amssymb}
\usepackage{mathtools}
\usepackage[version=4]{mhchem}
\usepackage{physics}
\usepackage{bigints}
\usepackage{setspace}
\usepackage{booktabs}
\usepackage{tabularx}
\usepackage{makecell}

\usepackage[super,sort&compress,numbers]{natbib}
\usepackage[colorlinks=true,allcolors=black]{hyperref}

\usepackage{xcolor}
\usepackage[normalem]{ulem}
\usepackage{changes}
\usepackage{todonotes}

\definecolor{addedcolor}{rgb}{0,0.5,0}
\definecolor{deletedcolor}{rgb}{1,0,0}
\setdeletedmarkup{\textcolor{deletedcolor}{\sout{#1}}}
\setaddedmarkup{\textcolor{addedcolor}{#1}}

\usepackage{newunicodechar}
\newunicodechar{​}{} 
\DeclareUnicodeCharacter{2009}{\,}
\DeclareUnicodeCharacter{2212}{\ensuremath{-}}

\title{\textbf{Amorphous silicon metasurfaces encode light-driven catalytic activity}}
\author{%
Elif Nur Dayi$^{1}$,
Priscila Vensaus$^{1}$,
Milad Sabzehparvar$^{1}$,
Michele Puppin$^{2}$,
Omer Can Karaman$^{1}$,
Giulia Tagliabue$^{1,*}$\\[0.75em]
{\small
\textsuperscript{1}Laboratory of Nanoscience for Energy Technologies (LNET), STI,\\
École Polytechnique Fédérale de Lausanne, 1015 Lausanne, Switzerland\\[0.75em]
\textsuperscript{2}Lausanne Centre for Ultrafast Science (LACUS), STI,\\
École Polytechnique Fédérale de Lausanne, 1015 Lausanne, Switzerland\\[0.75em]
\textsuperscript{*}Correspondence: giulia.tagliabue@epfl.ch
}%
}

\begin{document}
\setstretch{1.15}
\begin{titlepage}
\centering

\vspace*{1cm}

{\Large\bfseries Strong Directional Asymmetry through Modal Hybridization in All-Dielectric Metasurfaces}

\vspace{1cm}

{\normalsize
Elif Nur Dayi$^{1}$, Omer Can Karaman$^{1}$, Diana Dall'Aglio$^{1}, $ Giulia Tagliabue$^{1,*}$
\par}

\vspace{0.6cm}

{\small
$^{1}$Laboratory of Nanoscience for Energy Technologies (LNET), STI,\\
École Polytechnique Fédérale de Lausanne, 1015 Lausanne, Switzerland
\par}
\vspace{0.2cm}

{\small
$^{*}$Correspondence: giulia.tagliabue@epfl.ch
\par}

\vspace{1cm}

\vspace{1cm}

\begin{abstract}
\small
Direction-dependent optical responses enable multifunctional metadevices but often require spatially varying meta-atoms or precisely aligned multilayer architectures. Here, we demonstrate a simple amorphous-silicon nanodisk array that exhibits broadband directional spectral asymmetry under unpolarized illumination, without requiring complex patterning or stacking. Coupling between localized Mie resonances and nonlocal guided-mode resonances supported by the residual a-Si film produces hybrid resonances whose excitation and interference depend on the illumination direction. Upon reversal of the illumination direction, the reflection and absorption spectra are reshaped, while reciprocal transmission remains identical. At normal incidence, the measured absolute forward–backward reflectance difference reaches \(24.2\%\) at \(616~\mathrm{nm}\), with a mean absolute difference of \(8.3\%\) across \(500\text{–}850~\mathrm{nm}\). A locally optimized geometry yields a simulated absolute directional absorptance difference of \(79.7\%\) at \(800~\mathrm{nm}\) and a mean absolute difference of \(17.4\%\) over the same range. Angle-resolved measurements reveal additional spectral control through distinct dispersion sensitivity of resonances. We exploit these responses to encode patterns through local variations in nanodisk width, enabling spectral filtering and direction-dependent visibility. We further demonstrate a functional QR code that is readable in forward reflection and transmission but concealed in backward reflection. Fabricated in a single lithographic patterning step, this platform combines geometric simplicity with multiple optical readout channels for a vast range of applications including authentication, information encoding, and direction-selective light harvesting.
\end{abstract}

\end{titlepage}

\section{Introduction} \label{sec:Intro}

Conventional optical elements often require separate or cascaded components for different functions, increasing their footprint, alignment complexity, and optical losses. Integrating multiple optical functionalities within a single compact device has therefore become a growing focus in nanophotonics, as it reduces system complexity while enabling multiple optical responses or information channels within the same architecture \cite{kim_bidirectional_2024,chen_directional_2020,yazdi_bianisotropic_2015}. Alongside wavelength, polarization, and incidence angle, illumination direction provides an additional control parameter for accessing such multifunctional responses.

Metasurfaces, two-dimensional arrays of subwavelength resonators, provide a versatile platform for exploiting this directional degree of freedom. All-dielectric metasurfaces are particularly attractive because their strong electric and magnetic resonances enable control over the amplitude, phase, polarization, and wavefront of light within an ultrathin optical element. Generally referred to as Janus, bifacial, or bidirectional metasurfaces, these devices can produce distinct optical responses when illuminated from opposite sides \cite{ chen_directional_2020, chen_bifacial-metasurface-enabled_2022,ji_janus_2026,yang_direction-duplex_2023}. This direction-dependent behavior has enabled applications including holography, optical multiplexing, information encryption, displays, and direction-selective absorption \cite{chen_directional_2020,wang_asymmetric_2026,chang_directionally_2025,qiu_full-space_2026,kim_asymmetric_2020}.

Nonreciprocal reflection has been demonstrated using space–time-modulated metasurfaces, where optically driven temporal modulation enables direction-dependent frequency conversion \cite{guo_nonreciprocal_2019,efimov_nonreciprocal_2026}. Notably, a direction-dependent optical response does not necessarily require broken reciprocity. Lorentz reciprocity requires equal transmission between corresponding optical channels when the source and detector are interchanged in a linear, time-invariant system composed of reciprocal materials\cite{potton_reciprocity_2004}. At normal incidence, the corresponding channels are the forward and backward waves with the same polarization, resulting in identical power transmittance, whereas the reflectance can differ between opposite sides.

Directional asymmetry in reciprocal systems has been realized through various mechanisms, including structural asymmetry \cite{rodriguez_breaking_2014,butun_asymmetric_2015}, bianisotropy \cite{yazdi_bianisotropic_2015,hostalet-vicent_novel_2026}, and polarization conversion \cite{wu_asymmetric_2022,gao_reconfigurable_2025,fu_asymmetric_2022}. Yazdi et al. \cite{yazdi_bianisotropic_2015} showed asymmetric resonant absorption by combining nonidentical resonator layers on opposite sides of a thin dielectric slab. At optical frequencies, Kim et al. \cite{kim_asymmetric_2020} demonstrated direction-dependent reflective colours and solvent-tunable information concealment in an asymmetric metal--\allowbreak dielectric--\allowbreak metal Fabry–Perot cavity. Wu et al.~\cite{wu_asymmetric_2022} realized direction-dependent polarization conversion in hyperbolic quartz, where light incident from opposite sides coupled differently to orthogonal polarization channels due to the material anisotropy. Chiral metasurfaces can similarly produce asymmetric reflection or transmission through polarization conversion under circularly polarized illumination \cite{gao_reconfigurable_2025,chen_bifacial-metasurface-enabled_2022,yang_direction-duplex_2023}. Since this response depends on the incident polarization, these implementations require polarization control and their directional contrast can be reduced or eliminated under unpolarized illumination \cite{fu_asymmetric_2022}.
An alternative route to achieving a directional optical response is through modal interference. Yang and Chen numerically estimated asymmetric absorption by hybridizing dielectric Mie and plasmonic resonances in Si/Au resonator arrays \cite{yang_hybridization_2020}. In plasmonic nanoparticle arrays, selective coupling between localized surface plasmon resonances and collective lattice modes has also produced differences between forward and backward photoluminescence emission \cite{rodriguez_breaking_2014}. Similarly, Shevchenko et al. \cite{shevchenko_bifacial_2015} reported asymmetric reflection via multipolar excitations in gold nanodisc dimers embedded in glass, where a quadrupole mode dominated illumination from one side and a dipolar mode from the other at 675 nm. More recently, Manez-Espina et al. \cite{manez-espina_strong_2026} theoretically predicted dual-band directional absorption in the near-infrared through strong coupling between electric and magnetic quasi-bound states in the continuum in dielectric nanodisk metasurfaces.

Despite this progress, existing implementations often rely on multilayer, multimaterial, or cascaded architectures~\cite{yazdi_bianisotropic_2015,butun_asymmetric_2015,yang_hybridization_2020,yildirim_one-way_2020,chen_directional_2020}, or on spatially varying meta-atoms that encode direction-dependent phase profiles~\cite{chen_directional_2020,chang_directionally_2025}. Other approaches exploit chiral or polarization-conversion channels and therefore require control of the incident polarization or polarization-resolved detection~\cite{gao_reconfigurable_2025,fu_asymmetric_2022}. Many demonstrations are also optimized for discrete wavelengths or frequencies and prescribed illumination states~\cite{shevchenko_bifacial_2015,yildirim_one-way_2020,manez-espina_strong_2026,qiu_full-space_2026}, limiting their operation under broadband, unpolarized illumination. These limitations leave open the question of whether a simple, uniform dielectric array can experimentally generate pronounced broadband spectral asymmetry upon reversal of the illumination direction, including changes in spectral shape rather than only in overall magnitude.

Here, we report direction-selective responses arising from interference between Mie and guided-mode resonances in a periodic a-Si nanodisk array under broadband, unpolarized illumination. This monolithic platform combines a simple nanodisk geometry with a single lithographic patterning step, reducing fabrication complexity. The system is passive and does not rely on external bias, polarization control, or high-intensity excitation. Experimentally, the peak forward–backward reflectance difference reaches \(24.2\) percentage points at \(616~\mathrm{nm}\), with a mean absolute difference of \(8.3\) percentage points across \(500\text{–}850~\mathrm{nm}\). We use full-wave simulations and modal analysis to determine how the overlap and coupling between the resonances govern the directional optical response. Parametric studies identify geometries offering either strong directional contrast or near-unity forward absorption. One reaches a simulated absorptance difference of \(79.7\) percentage points at \(800~\mathrm{nm}\) and a mean absolute contrast of \(17.4\) percentage points across \(500\text{--}850~\mathrm{nm}\), while another achieves \(99.8\%\) forward absorptance at \(702~\mathrm{nm}\). Simulations using TiO$_2$, CH$_3$NH$_3$PbI$_3$ (MAPI), and Ge further indicate that the underlying mechanism can be extended across different material platforms and spectral regions, from the ultraviolet to the near-infrared.

The nonlocal guided-mode resonances provide an additional means of controlling the optical response because their spectral positions depend strongly on the in-plane momentum of the incident light. This intrinsic angular dispersion has previously been exploited for angle-selective sensing and wavefront control \cite{leitis_angle-multiplexed_2019,kamali_angle-multiplexed_2017}. More recently, Cohen \textit{et al.}\cite{cohen_engineering_2026} engineered angle-dependent transmission by hybridizing localized nanopillar resonances with nonlocal lattice-mediated interactions. However, these studies have focused on angular selectivity without examining how it can coexist with pronounced front-to-back reflection asymmetry in the same dielectric metasurface. We therefore map the evolution of the hybridized modes with incidence angle, showing that illumination direction and incidence angle provide complementary control over the spectral response. Since the angular dependence arises from the intrinsic dispersion of the resonances, it does not require spatially encoded phase profiles or complex unit cells.

Lastly, we exploit the direction-dependent response to encode optical patterns through local variations in nanodisk width. A QR-code-like pattern can be selectively concealed or revealed by combining illumination direction with spectral filtering, while a functional meta-QR code enables direction- and channel-selective machine readout, with successful decoding in forward reflection and transmission and suppressed readability in backward reflection under matched acquisition conditions. The encoded features are observable under broadband, unpolarized illumination without wavelength-selective laser excitation. Exposure controls and grayscale analysis confirm that the directional contrast persists across different integration times and without chromatic information. Quantitative intensity analysis shows an approximately \(83\%\) reduction in the Michelson contrast magnitude upon reversing the reflection direction, while transmission provides an additional readout channel. These results establish illumination direction, detection channel, and spectral filtering as complementary parameters for optical information multiplexing within a single patterned layer, offering opportunities for optical authentication, anti-counterfeiting, and view-dependent information displays.

\section{Directional optical response of the all-dielectric metasurface}

\begin{figure}[H]
    \centering  
\includegraphics[width=17cm]{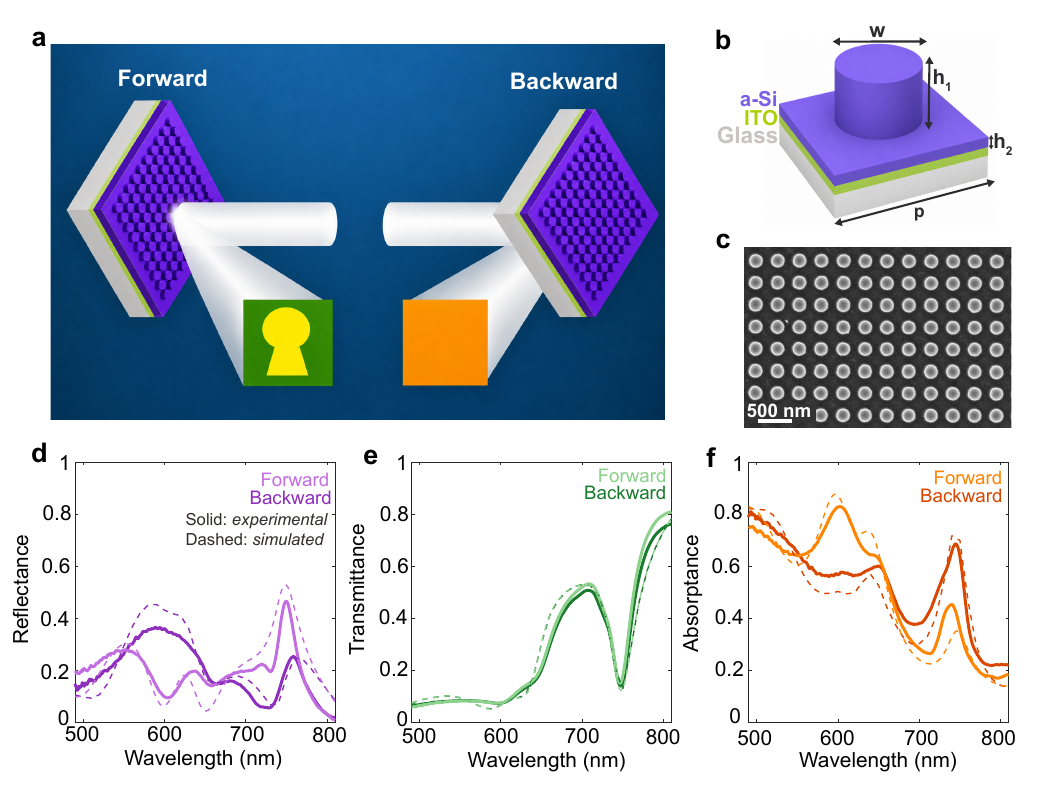}
    \caption{\textbf{Directional optical response of the a-Si directional metasurface.} \textbf{a}, Conceptual illustration of the metasurface under collimated, near-normal illumination from the forward and backward directions, producing distinct reflected optical outputs. \textbf{b}, Schematic of the unit cell, consisting of an a-Si nanodisk and residual a-Si layer on ITO-coated fused silica. Here, $w$ is the nanodisk diameter, $h_1$ is the nanodisk height, $h_2$ is the residual a-Si thickness, and $p$ is the array periodicity. \textbf{c}, Top-view SEM image of the fabricated nanodisk array. Scale bar: $500~\mathrm{nm}$. \textbf{d--f}, Experimental and simulated \textbf{d} reflectance, \textbf{e}, transmittance, and \textbf{f}, absorptance spectra under forward and backward illumination for a metasurface with $w=130~\mathrm{nm}$, $h_1=110~\mathrm{nm}$, $h_2=120~\mathrm{nm}$, and $p=325~\mathrm{nm}$. Solid and dashed curves represent experimental measurements and full-wave simulations, respectively. Absorptance was calculated as $A=1-R-T$.}
    \label{fig-schematic}
\end{figure}

The direction-selective response of the metasurface is illustrated in Figure~\ref{fig-schematic}a. We define forward illumination as incidence from the patterned metasurface side and backward illumination as incidence through the substrate. Reversing the illumination direction produces different reflection spectra and, consequently, asymmetric absorption across the visible and near-infrared spectral range. Consistent with reciprocity, transmission remains identical for the two directions. The same structure therefore provides three distinct optical responses without external bias or reconfiguration.

Figure~\ref{fig-schematic}b shows a representative meta-atom and the geometric parameters used to tune the supported resonances. The metasurface consists of a-Si nanodisks with width $w$ and height $h_1$ protruding from a residual a-Si film of thickness $h_2$ deposited on an ITO-coated fused-silica substrate. The nanodisks are arranged in a square lattice with periodicity $p$. Together, $w$, $h_1$, $h_2$, and $p$ determine the spectral positions and coupling of the resonances. A top-view scanning electron micrograph of the fabricated array is shown in Figure~\ref{fig-schematic}c. The fabrication procedure is described in \nameref{sec:Methods} and illustrated in Supporting Figure~1.

Figures~\ref{fig-schematic}d--f compare the experimental and simulated reflectance, transmittance, and absorptance spectra for a representative metasurface with $w=130~\mathrm{nm}$, $h_1=110~\mathrm{nm}$, $h_2=120~\mathrm{nm}$, and $p=325~\mathrm{nm}$. Purple, green, and orange curves represent reflectance, transmittance, and absorptance, respectively. Within each color family, darker and lighter shades indicate backward and forward illumination, while solid and dashed curves represent experiment and simulation. Reflectance and transmittance were measured directly, and absorptance was calculated as $A=1-R-T$.

Full-wave electromagnetic simulations (COMSOL Multiphysics) were performed using the wavelength-dependent optical constants of a-Si and ITO measured by spectroscopic ellipsometry and presented in Supporting Figure~2. Further details of the numerical simulations and optical measurements are described in ~\nameref{sec:Methods}. The simulations reproduce the measured resonance positions and overall spectral line shapes under both illumination directions. In particular, the model captures the identical transmission spectra together with the pronounced directional differences in reflection and absorption, providing a validated basis for the subsequent modal analysis.

Since the fabricated structure contains a transparent conductive oxide underlayer, we independently assessed its contribution to the directional response. ITO was included to facilitate integration with optoelectronic and energy-conversion systems but is not essential to the underlying optical mechanism. Simulations without ITO preserve the principal resonant features and forward--backward absorptance contrast, with only minor changes in the visible response (Supporting Figure~3).

Crucially, the directional asymmetry observed in Figure~\ref{fig-schematic}d is not confined to a single operating wavelength. Reversing the illumination direction redistributes the relative strengths of multiple peaks and dips, changing the reflectance line shape rather than producing a uniform intensity offset. We quantify this broadband response using the mean absolute reflectance difference,
\begin{equation}
\left\langle |\Delta R| \right\rangle =
\frac{1}{N}\sum_{i=1}^{N}
\left|R_{\mathrm{F}}(\lambda_i)-R_{\mathrm{B}}(\lambda_i)\right|,
\end{equation}
where the spectrum is sampled at $N$ uniformly spaced wavelengths. The measured absolute reflectance difference reaches $24.2$ percentage points at $616~\mathrm{nm}$, while $\left\langle |\Delta R| \right\rangle=8.3$ percentage points across the full $500$--$850~\mathrm{nm}$ range, demonstrating that the response extends over a $350~\mathrm{nm}$ bandwidth. The same definition can be applied to absorptance by replacing $R$ with $A$. This wavelength-dependent spectral effect points to direction-dependent excitation and interference among multiple resonant modes. Upon demonstrating the broadband character of the asymmetry and its persistence without ITO, we next use the full-wave electromagnetic model validated against the measured spectra in Figure~\ref{fig-schematic}d--f to identify the contributing modes, determine how their coupling produces the directional response and explore a range of design parameters to enhance directional contrast.

\section{Hybridization of Mie and Guided-mode Resonances}
\label{sec:modal_analysis}

\begin{figure}[H]
    \centering  
\includegraphics[width=18cm]{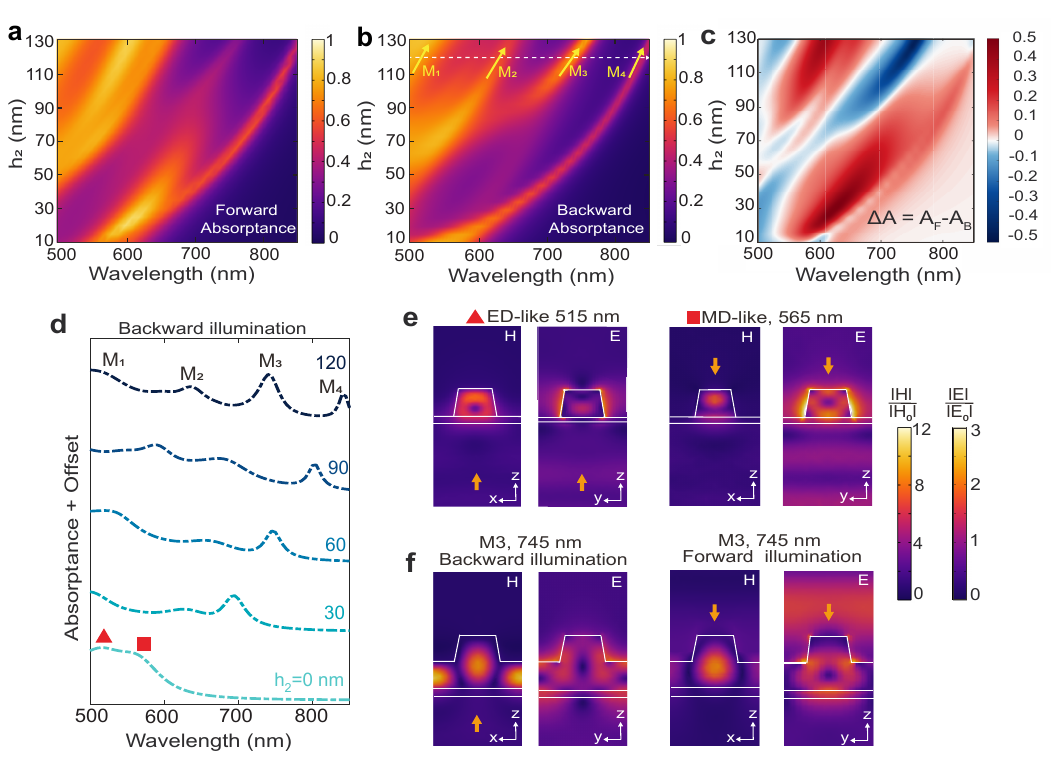}
    \caption{\textbf{Modal evolution and direction-dependent field excitation in the metasurface.} \textbf{a,b}, Simulated absorptance under forward and backward illumination, respectively, as a function of wavelength and residual-film thickness $h_2$, with $w=130~\mathrm{nm}$, $h_1=110~\mathrm{nm}$, and $p=325~\mathrm{nm}$. The dashed line in \textbf{b} marks the representative structure with $h_2=120~\mathrm{nm}$, and the arrows identify the spectral features $M_1$--$M_4$. \textbf{c}, Directional absorptance difference, defined as $\Delta A=A_{\mathrm{F}}-A_{\mathrm{B}}$. Positive and negative values indicate stronger absorptance under forward and backward illumination, respectively. \textbf{d}, Vertically offset backward-illumination spectra for selected residual-film thicknesses from $h_2=0$ to $120~\mathrm{nm}$, illustrating the progressive restructuring of the resonances as the residual a-Si film is introduced. The triangle and square mark the ED-like and MD-like resonances of the $h_2=0$ nanodisk array. \textbf{e}, Normalized magnetic-field magnitude in the $xz$ plane and electric-field magnitude in the $yz$ plane for the ED-like and MD-like resonances at $515$ and $565~\mathrm{nm}$, respectively. \textbf{f}, Corresponding normalized field distributions of the $M_3$ feature at $745~\mathrm{nm}$ under backward and forward illumination for the complete structure with $h_2=120~\mathrm{nm}$. Yellow arrows indicate the illumination direction. The field quantities are normalized to their respective incident-field amplitudes, and corresponding maps are displayed using common color scales. The simulated geometry incorporates a sidewall taper of 30 nm.}

    \label{abs-device}
\end{figure}

To identify the modal origin of the directional response, we vary the residual-film thickness $h_2$ from zero, corresponding to isolated nanodisks, to the value used in the complete device. We focus on absorptance to track the resonant branches, which can manifest as either peaks or dips in reflectance and transmittance. Figure~\ref{abs-device}a,b map the simulated absorptance as a function of wavelength and $h_2$ under forward and backward illumination. The nanodisk diameter, height, and periodicity are fixed at $w=130~\mathrm{nm}$, $h_1=110~\mathrm{nm}$, and $p=325~\mathrm{nm}$, with $h_2=120~\mathrm{nm}$, matching the device characterized in Figure~\ref{fig-schematic}. Similar spectral branches appear for both directions but with markedly different intensities, indicating direction-dependent coupling to the same family of modes. The resulting contrast, $\Delta A=A_{\mathrm{F}}-A_{\mathrm{B}}$, is mapped in Figure~\ref{abs-device}c. Alternating positive and negative bands track the resonant branches and shift systematically with $h_2$, showing that the residual film controls both the resonance wavelengths and the favored illumination direction.

With equal transmission in both directions, $\Delta A=-\Delta R$, corresponding to a wavelength-dependent redistribution between reflection and absorption.
To isolate the contribution of thin-film interference, we simulated planar a-Si films with thicknesses matching the residual and total a-Si layers (Supporting Figure~4). Direction reversal primarily changes the response magnitude, whereas varying the film thickness shifts the broad interference envelope. Thin-film interference therefore contributes a directional background but cannot reproduce the alternating contrast bands and sign reversals arising from the resonances of the patterned structure.

Figure~\ref{abs-device}d presents selected backward-illumination spectra as $h_2$ is varied from $0$ to $120~\mathrm{nm}$. Without the residual film, two prominent resonances appear at approximately $515$ and $565~\mathrm{nm}$. Their field distributions in Figure~\ref{abs-device}e exhibit electric-dipole-like and magnetic-dipole-like character, respectively, consistent with localized Mie-type resonances of the nanodisks \cite{spinelli_broadband_2012}. No pronounced near-infrared resonance is observed for $h_2=0$, and the absorptance decreases rapidly toward longer wavelengths. Introducing the residual a-Si film produces additional branches and substantial spectral restructuring rather than a uniform shift of the original resonances. At $h_2=120~\mathrm{nm}$, four spectral features, denoted $M_1$--$M_4$, can be distinguished. Their evolution with $h_2$ is consistent with coupling between the localized nanodisk resonances and guided-mode responses of the underlying a-Si film, producing hybrid modes of the metasurface.

Cross-sectional high-angle annular dark-field scanning transmission electron microscopy (HAADF-STEM) revealed tapered sidewalls in the fabricated nanodisks, as shown in Supporting Figure~5a. Throughout this work, the nanodisk width $w$ denotes the top diameter, while the taper $t$ is defined as the radial increase toward the bottom, such that $w_{\mathrm{bottom}}=w+2t$. A taper of $t=30~\mathrm{nm}$ was estimated from the cross sections and used for all simulated nanodisk widths. Because taper breaks the out-of-plane mirror symmetry of the individual resonators, it can contribute to the directional response. Rodriguez \textit{et al.}~\cite{rodriguez_breaking_2014} reported asymmetric forward and backward luminescence from tapered plasmonic arrays due to direction-dependent coupling between different lattice resonances. We therefore compared the tapered structure with a control containing cylindrical nanodisks with no taper. The cylindrical structure exhibits asymmetric reflection and absorption together with identical transmission (Supporting Figure~5b--d), demonstrating that taper is not required for the directional response. Increasing $t$ from $0$ to $60~\mathrm{nm}$ shifts the resonances and modifies their relative strengths (Supporting Figure~5e), while the direction-dependent field distributions at $M_3$ persist for $t=0$ (Supporting Figure~5f,g). Taper therefore tunes the detailed spectral response but does not generate the underlying asymmetry, which arises from direction-dependent excitation and interference among the localized nanodisk, guided-mode resonances and the contribution of the overall stack.

Figure~\ref{abs-device}f shows the normalized field distributions of the $M_3$ feature at $745~\mathrm{nm}$ for the complete structure with $h_2=120~\mathrm{nm}$. Reversing the illumination direction redistributes the field between the nanodisk and residual a-Si film. The interfacial field concentration and extended modal profile indicate hybridization between localized nanodisk and guided-film contributions, while the angular dispersion examined in the following section provides further evidence of the GMR character of $M_3$.

To assess the broader structural tunability of the directional response, we extended the analysis beyond $h_2$ through independent sweeps of the nanodisk width $w$, height $h_1$, and periodicity $p$. The resulting absorptance maps show that $w$ and $h_1$ primarily shift the resonances, whereas $p$ and $h_2$ produce stronger changes in modal overlap and directional preference (Supporting Figure~6). The maximum forward- and backward-favoring contrasts obtained from these sweeps are summarized in Supporting Figure~7 and discussed in Supporting Note~1. The single-parameter sweeps reveal how each geometric parameter influences the directional response, but the parameters cannot be optimized independently because they jointly shift and hybridize the resonances. A geometry combining values selected from the independent sweeps reaches a peak absorptance difference of $\Delta A=0.361$ and a mean absolute spectral contrast of $7.9$ percentage points across $500$--$850~\mathrm{nm}$. Joint optimization increases these values to $\Delta A=0.797$ at $800~\mathrm{nm}$ and a mean absolute spectral contrast of $17.4$ percentage points (Supporting Figure~8a,b). A separate design targeting high forward absorption reaches $A_{\mathrm{F}}=0.998$ at $702~\mathrm{nm}$ while maintaining $\Delta A=0.723$ (Supporting Figure~8c). These results show that the geometry can be tailored toward either maximum directional contrast or near-unity one-sided absorption, with potential applications in direction-selective photodetection, photochemical interfaces, energy conversion, and antireflection coatings\cite{spinelli_broadband_2012,liu_multipole_2020,dayi_metaphotonic_2026}.

To test whether this behavior extends beyond a-Si, we simulated asymmetric metasurfaces composed of TiO$_2$, CH$_3$NH$_3$PbI$_3$ (MAPI), and Ge. The wavelength-dependent optical constants were taken from Sarkar \textit{et al.}~\cite{sarkar_hybridizedguided-mode_2019} for TiO$_2$, Phillips \textit{et al.}~\cite{phillips_dispersion_2015} for CH$_3$NH$_3$PbI$_3$, and Amotchkina \textit{et al.}~\cite{amotchkina_characterization_2020} for Ge. These results indicate that the underlying mechanism is transferable across material platforms, although the geometry must be optimized for each material and target spectral range.

\section{Angle-dependent spectral control}
\label{sec:angle_dependence}

\begin{figure}[H]
    \centering  
\includegraphics[width=17cm]{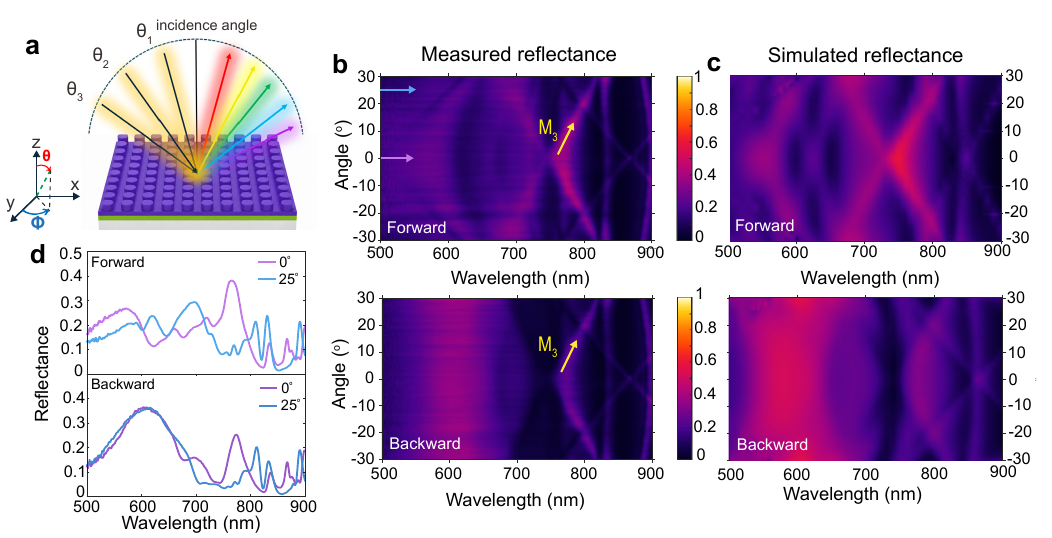}
    \caption{\textbf{Angle-dependent directional reflection.} \textbf{a}, Schematic illustration of angle-dependent illumination, defining the polar incidence angle $\theta$ and azimuthal angle $\phi$. \textbf{b}, Experimentally measured reflectance as a function of wavelength and incidence angle under forward (top) and backward (bottom) illumination. All measurements and simulations correspond to the representative metasurface analyzed in Figures~\ref{fig-schematic} and~\ref{abs-device}, with $w=130~\mathrm{nm}$, $h_1=110~\mathrm{nm}$, $h_2=120~\mathrm{nm}$, and $p=325~\mathrm{nm}$.  The yellow arrows highlight the angularly dispersive branch associated with the hybrid $M_3$ resonance identified in Figure~\ref{abs-device}. The purple and blue arrows indicate the spectra extracted at $\theta=0^{\circ}$ and $25^{\circ}$, respectively. \textbf{c}, Reflectance maps obtained from full-wave electromagnetic simulations for forward (top) and backward (bottom) illumination. The simulated positive-angle response is mirrored about $\theta=0^{\circ}$ according to the in-plane symmetry of the metasurface. \textbf{d}, Measured forward and backward reflectance spectra at $\theta=0^{\circ}$ and $25^{\circ}$, showing the angular redistribution of the resonant response.}
\label{fig:angle-dependence}
\end{figure}

To further distinguish the localized and lattice-mediated contributions, we exploit their different sensitivities to the in-plane momentum. Guided-mode resonances are strongly angle dispersive, whereas localized Mie-type and Fabry--Perot-like responses vary more weakly with incidence angle. We therefore examine the angle-resolved optical response to trace the dispersive branches and assess incidence angle as an additional tuning parameter. We performed angle-dependent reflectance measurements on the representative metasurface investigated in Figures~\ref{fig-schematic} and~\ref{abs-device}, with $w=130~\mathrm{nm}$, $h_1=110~\mathrm{nm}$, $h_2=120~\mathrm{nm}$, and $p=325~\mathrm{nm}$. Measurements were performed under unpolarized illumination. The polar angle of incidence, $\theta$, was varied as illustrated in Figure~\ref{fig:angle-dependence}a.
The forward- and backward-reflection maps in Figure~\ref{fig:angle-dependence}b reveal two distinct regimes of angular dispersion. The broad spectral features below approximately $700~\mathrm{nm}$ shift only weakly with angle, whereas the near-infrared resonances undergo pronounced shifts and splitting. In particular, the $M_3$ feature identified near $745~\mathrm{nm}$ at normal incidence develops into a strongly dispersive branch, supporting its assignment as a hybrid mode with a substantial guided-mode contribution.

The simulated reflectance maps reproduce the measured branch positions and their evolution with angle under both illumination directions (Figure~\ref{fig:angle-dependence}c). At nonzero polar incidence angles, the response can depend on both polarization and the azimuthal orientation of the in-plane wavevector relative to the square lattice. We therefore calculated the TE and TM responses for propagation along two primary directions along the lattice axis ($\phi=0^{\circ}$) and the lattice diagonal ($\phi=45^{\circ}$). The four responses were equally averaged to approximate the unpolarized illumination used experimentally, while the individual polarization- and azimuth-resolved maps are presented in Supporting Figure~10. The X-shaped dispersion is consistent with the lifting of the normal-incidence degeneracy between oppositely propagating guided modes as the in-plane wavevector increases. By contrast, the localized Mie-type resonances exhibit weaker angular dispersion because they do not rely on in-plane propagation. Varying $\theta$ therefore changes the spectral overlap between these weakly dispersive contributions and the guided-mode resonance, modifying their hybridization and interference. Switching the illumination direction leaves the resonant branch positions nearly unchanged but alters their spectral weights, showing that forward and backward illumination couple differently to the same hybrid modes.

Figure~\ref{fig:angle-dependence}d compares the measured spectra at incidence angles of $\theta=0^{\circ}$ and $25^{\circ}$. Changing the incidence angle strongly restructures the near-infrared response while producing smaller changes in the broader visible-wavelength features. The angular evolution also differs between forward and backward illumination, causing the directional contrast to vary continuously with wavelength and angle. Across the measured angular range, the absolute forward--backward reflectance difference, $|\Delta R|=|R_{\mathrm{F}}-R_{\mathrm{B}}|$, reaches $27.3$ percentage points at $\lambda=716~\mathrm{nm}$ and $\theta=-17^{\circ}$. The mean absolute difference across the measured spectral range of $500$--$850~\mathrm{nm}$ reaches $9.9$ percentage points at $\theta=-28^{\circ}$. Illumination direction and incidence angle therefore provide complementary control over the spectral response.

Angle-dependent transmission measurements are presented in Supporting Figure~11. These measurements  cover an incidence-angle range of approximately $-12^{\circ}$ to $+12^{\circ}$ because a different objective was used to improve the angular sampling of the back focal plane to preserve resolution. Although the transmission spectrum evolves with incidence angle, it remains equivalent under forward and backward illumination within the experimental variation, consistent with reciprocity and the simulated response.

To isolate the role of the residual a-Si film in the angular response, we also examined a control metasurface without a continuous residual layer ($h_2\approx 0$), shown in Supporting Figure~12. The measured maps reveal a weak, broad angle-dependent feature near $600$--$700~\mathrm{nm}$, particularly under backward illumination. This feature is consistent with the localized nanodisk resonances identified for $h_2=0$, whose spectral response can still be modified by the incidence angle and the asymmetric surrounding media. However, the pronounced dispersive branches observed in the partially etched metasurface are absent in this case, supporting their assignment to guided-mode resonances sustained by the residual layer. The control sample also shows a forward--backward reflectance difference because the nanodisks are exposed to air on one side and supported by ITO-coated fused silica on the other. These results indicate that the residual film enables the strongly dispersive guided-mode contribution and its hybridization with the localized resonances, but is not solely responsible for the underlying directional asymmetry.

Collectively, these results show that the optical response is not limited to two discrete forward and backward states. Illumination direction determines the coupling strength to the hybrid modes, while incidence angle continuously shifts the guided-mode contribution relative to the less dispersive localized resonances. The metasurface therefore combines directional asymmetry with continuous angular spectral tunability, without active modulation or spatially varying phase profiles. This response could support angle-sensitive filters, view-dependent optical elements, sensing, and illumination-selective absorption. In the following section, we exploit the viewpoint-dependent spectral response to encode and selectively reveal optical information.

\section{Direction-selective optical information encoding}
\label{sec:information_multiplexing}

Resonant dielectric metasurfaces translate nanoscale variations in resonator geometry into changes in spectral response and perceived color, as widely exploited in structural-color and resonant-absorber platforms \cite{yang_direction-duplex_2023,england_optical_2017}. Here, we combine spatial geometric encoding with directional and spectral selectivity. The encoded regions are defined solely through local variations in nanodisk width, while the layer thicknesses and lattice arrangement remain unchanged within each design. The resulting differences in reflection and transmission provide spatial contrast that can be controlled through the illumination direction, detection channel, and spectral range.

\begin{figure}[H]
    \centering  
\includegraphics[width=17cm]{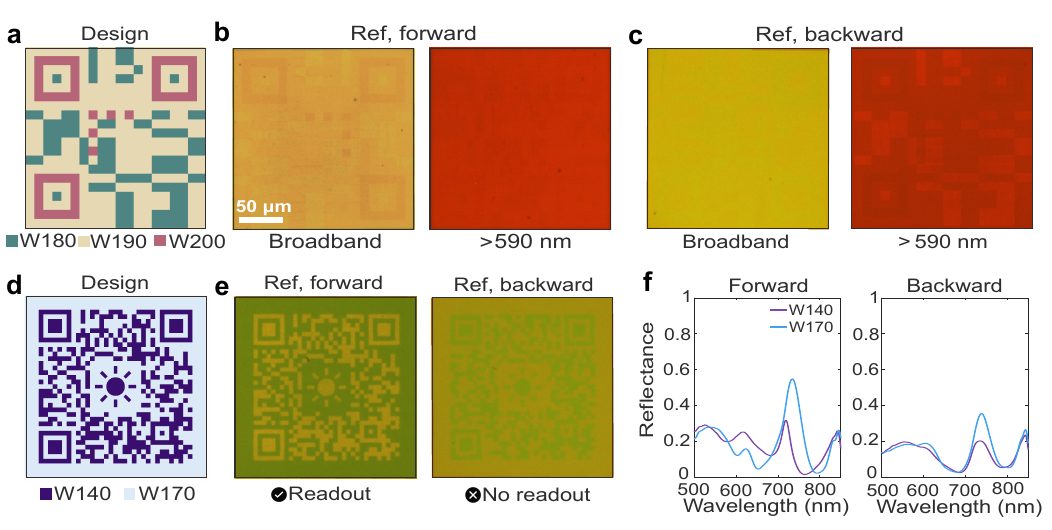}
    \caption{\textbf{Direction- and wavelength-selective readout of encoded metasurface patterns.}
\textbf{a}, Design of a QR-code-like pattern encoded using nanodisk widths of $w=180$, 190, and 200~nm with a periodicity of $p=380$~nm.
\textbf{b,c}, Reflection micrographs recorded under forward and backward illumination, respectively, using broadband illumination and a 590~nm long-pass filter. Within each forward--backward comparison, the illumination and acquisition conditions were kept identical, and the illumination direction was reversed solely by physically flipping the sample. Spectral filtering suppresses the pattern under forward illumination while revealing it under backward illumination.
\textbf{d}, Design of a functional QR code encoded using nanodisk widths of $w=140$ and $170~\mathrm{nm}$ with $p=325~\mathrm{nm}$.
\textbf{e}, Forward and backward reflection micrographs acquired with an exposure time of 75~ms, detector gain of 1, and a microscope-lamp power setting of 40\%. The sample was physically flipped without changing these settings. QR-code readout succeeds under forward illumination but fails under backward illumination.
\textbf{f}, Mirror-normalized reflectance spectra of the $w=140$ and 170~nm regions under forward and backward illumination. The greater spectral separation between the two regions under forward illumination supports the stronger image contrast and successful QR-code readout observed in \textbf{e}; their reduced separation under backward illumination results in lower contrast and readout failure. The $50~\mu\mathrm{m}$ scale bar applies to all optical micrographs in the first row (b,c). The meta-QR code in the second row occupies a $200\times200~\mu\mathrm{m}^{2}$ area.}
\label{fig:patterns}
\end{figure}

Figures~\ref{fig:patterns}a--c demonstrate the combined use of illumination direction and spectral filtering to control pattern visibility. Forward and backward reflection images were acquired under identical conditions within each comparison. The sample was physically flipped while the illumination level, objective, exposure time, detector gain, and other imaging settings were kept constant. Thus, the contrast changes observed upon sample reversal arise from the direction-dependent optical response rather than differences in image acquisition.

 The QR-code-like pattern is encoded using three nanodisk widths, $w=180$, $190$, and $200~\mathrm{nm}$, with a fixed periodicity of $p=380~\mathrm{nm}$. Under broadband illumination, the different widths produce a distinct distribution of reflected intensities. Introducing a $590~\mathrm{nm}$ long-pass filter suppresses the pattern contrast under forward illumination while revealing it under backward illumination. The same static structure can therefore produce different optical readout states through the combined selection of illumination direction and spectral range.

We next investigated whether the directional contrast was sufficient to control machine-readable information for enabling directional devices. Figure~\ref{fig:patterns}d shows a functional QR code, hereafter referred to as a meta-QR, encoded using nanodisk widths of $w=140$ and $170~\mathrm{nm}$ with $p=325~\mathrm{nm}$. The acquired meta-QR image is successfully decoded using smartphone cameras under forward reflection but becomes unreadable after the sample is flipped, despite identical acquisition conditions (Figure~\ref{fig:patterns}e; see Supplementary Information for imaging conditions and readout tests).The reflectance spectra in Figure~\ref{fig:patterns}f demonstrate that the resonance peaks associated with the two nanodisk widths are spectrally more distinct under forward illumination, whereas their responses become more similar under backward illumination, reducing the spatial contrast required for successful readout.

Supporting Figure~13 demonstrates that information concealed in backward reflection can be recovered by switching the detection channel to transmission, which restores the pattern contrast and enables readout of the same meta-QR presented in Figure\ref{fig:patterns} d-e. Increasing the exposure time from $50$ to $125~\mathrm{ms}$ enhances the detected intensity without restoring readout, which shows that the loss of backward-reflection readability does not stem from insufficient image brightness. To quantify the intensity separation between the code and background, we use the signed Michelson contrast, $C_{\mathrm{M}}=(I_{\mathrm{code}}-I_{\mathrm{background}})/(I_{\mathrm{code}}+I_{\mathrm{background}})$, where the intensities are averaged over the respective regions \cite{pelli_measuring_2013}. Its magnitude measures their relative intensity contrast, while the sign identifies the brighter region. The contrast changes from $-0.104$ in forward reflection to $+0.018$ in backward reflection, corresponding to an approximately $83\%$ reduction in magnitude and a weak reversal of relative brightness. In transmission, $C_{\mathrm{M}}=-0.147$, consistent with the recovered pattern visibility and successful readout. Finally, grayscale conversion preserves the difference in visibility between forward and backward reflection, supporting intensity-based readout without requiring color discrimination.

These demonstrations show that information can be distributed across several linked parameters. Nanodisk geometry defines the encoded pattern, illumination direction selects the reflection state, spectral filtering modifies its visibility, and reflection or transmission provides an additional detection channel. This combination of multichannel optical decoding and geometric simplicity provides a scalable basis for authentication labels, anti-counterfeiting elements, and information displays with selectively accessible states \cite{mousavi_khaleghi_high_2025,zhou_holographically_2025}.

\label{sec:Outlook}
\section{Outlook}

We demonstrate a polarization-independent all-dielectric metasurface that produces pronounced forward--backward differences in reflection and absorption while preserving reciprocal transmission. At normal incidence, the measured reflectance difference reaches $24.2$ percentage points at $616~\mathrm{nm}$, with a mean absolute difference of $8.3$ percentage points across $500$--$850~\mathrm{nm}$. The response arises from direction-dependent excitation and interference between localized Mie resonances and film- and lattice-mediated guided modes. Coupling of these modes through the residual a-Si film reshapes the spectra upon reversal of the illumination direction, while their angular dispersion provides continuous tunability beyond the two directional states.

The ability to redistribute incident light between reflection and absorption can be advantageous for many applications. Multidimensional optimization produces a simulated directional absorptance difference approaching $0.80$, while a separate design achieves near-unity absorption under forward illumination. Such behavior could benefit direction-sensitive photodetectors, antireflection coatings, sensing platforms, ultrathin optical elements and energy-conversion surfaces whose operation depends on the direction or angular distribution of incident light\cite{grayli_near-unity_2025,liu_multipole_2020,spinelli_broadband_2012,wang_polarization-controlled_2023}. Simulations using different materials such as TiO$_2$, CH$_3$NH$_3$PbI$_3$ and Ge suggest that the underlying mechanism can be extended across different material platforms and spectral regions, from the ultraviolet to the infrared. 

We further translate the spectral asymmetry into optical information encoding by varying only the nanodisk width within otherwise uniform arrays. A QR-code-like pattern can be concealed or revealed through illumination direction and spectral filtering, while a functional meta-QR code supports machine readout in forward reflection and transmission but is concealed in backward reflection. Direction reversal reduces the Michelson contrast magnitude by approximately $83\%$. Illumination direction, spectral range, and detection channel therefore provide complementary readout parameters within a static device fabricated using a single lithographic patterning step. The combination of broadband operation, geometric simplicity, and material versatility establishes hybrid local--nonlocal mode engineering in vertically asymmetric dielectric metasurfaces as a route toward multifunctional optical devices.

\section{Methods} \label{sec:Methods}

\subsection*{Numerical simulations}

Full-wave electromagnetic simulations were performed in the frequency domain using the Wave Optics Module in COMSOL Multiphysics. The computational model comprised a three-dimensional periodic unit cell containing an a-Si nanodisk connected to a residual a-Si film on an ITO-coated fused-silica substrate. Unless otherwise specified, the experimentally estimated $30~\mathrm{nm}$ sidewall taper was included in the nanodisk geometry. The wavelength-dependent complex refractive indices of a-Si and ITO were determined by spectroscopic ellipsometry and incorporated directly into the model (Supporting Figure~2).

Forward and backward illumination corresponded to excitation from the a-Si nanodisk and fused-silica sides, respectively. Floquet periodic boundary conditions were applied along the in-plane directions to represent an extended square array, while perfectly matched layers terminated the computational domain along the out-of-plane direction. Plane-wave excitation was introduced through periodic ports over the spectral ranges shown in the corresponding figures. Reflectance and transmittance were calculated by summing the power carried by all propagating diffraction channels and normalizing it to the incident power. Diffraction orders with in-plane indices $m,n=-2,\ldots,+2$ were included at both ports.

For angle-dependent simulations, the polar angle $\theta$ was defined relative to the surface normal, while the azimuthal angle $\phi$ specified the orientation of the in-plane wavevector relative to a lattice axis. The unit-cell geometry was fixed while $\theta$ was varied. To approximate the unpolarized experimental illumination, we calculated the TE and TM responses for propagation along a lattice axis ($\phi=0^{\circ}$) and along the lattice diagonal ($\phi=45^{\circ}$) and assigned equal weight to all four power responses. These two azimuthal orientations sample the inequivalent propagation directions of the square lattice. The individual polarization- and azimuth-resolved reflectance maps are presented in Supporting Figure~10.

\subsection*{Nanofabrication}
Metasurfaces were fabricated on $520~\mu\mathrm{m}$-thick fused-silica substrates with lateral dimensions of $2~\mathrm{cm}\times2~\mathrm{cm}$, as illustrated in Supporting Figure~1. A $30~\mathrm{nm}$ ITO layer was deposited by radio-frequency sputtering (Pfeiffer SPIDER 600), followed by a nominally $230~\mathrm{nm}$-thick a-Si layer deposited at room temperature (Alliance-Concept DP 650).
A $120~\mathrm{nm}$-thick layer of ZEP 520A electron-beam resist was spin-coated onto the sample, followed by an Electra 92 conductive polymer overlayer (Zeon Chemicals) to mitigate charging during exposure. The nanodisk arrays were defined by electron-beam lithography using a Raith EBPG5000+ operated at $100~\mathrm{kV}$, with exposure doses ranging from $140$ to $240~\mu\mathrm{C}\,\mathrm{cm}^{-2}$. After development, the resist pattern served as an etch mask for transfer into the a-Si layer. Anisotropic reactive-ion etching was performed using an Adixen AMS200 at $0\,^{\circ}\mathrm{C}$. The etching duration was varied between $65$ and $95~\mathrm{s}$ across the fabricated samples to control the nanodisk height and residual a-Si film thickness. The three samples discussed in the main text were etched for $75~\mathrm{s}$, producing partially etched nanodisks connected by a continuous residual a-Si film. The fully etched control sample presented in the Supporting Figure 12 was etched for $95~\mathrm{s}$, removing the continuous residual a-Si layer. The remaining resist was removed by immersion in acetone, followed by microwave oxygen-plasma cleaning (Tepla 300; $500~\mathrm{W}$, an O$_2$ flow rate of $400~\mathrm{mL}\,\mathrm{min}^{-1}$, and a duration of $30~\mathrm{s}$).

\subsection*{Metasurface characterization}

The fabricated metasurfaces were characterized by scanning electron microscopy (SEM; Zeiss Crossbeam and Merlin) and atomic force microscopy (AFM; Bruker FastScan). The wavelength-dependent refractive index and extinction coefficient of the a-Si and ITO layers were determined by spectroscopic ellipsometry (Woollam RC2) and used as material inputs for the electromagnetic simulations.

Microscale reflectance and transmittance measurements were performed using an NT\&C setup consisting of an inverted microscope (Nikon Eclipse Ti2) coupled to a spectrometer (Princeton Instruments SpectraPro HRS-500), as described in our previous work~\cite{dayi_large-area_2025}. The sample was illuminated from below using an unpolarized laser-driven white-light source (Energetiq LDLS) coupled through an optical fiber with a $100~\mu\mathrm{m}$ core. No polarization-selective elements were introduced into the illumination path. For transmission measurements, the incident beam was collimated using a condenser, and the measured spectra were referenced to a bare fused-silica substrate. For near-normal reflectance measurements, the incident beam was focused at the center of the back focal plane of a $60\times$ objective (Nikon S Plan Fluor ELWD, NA = 0.7), producing collimated illumination normal to the sample. Reflectance was calibrated against a protected silver mirror with a known spectral response (Thorlabs PF10-03-P01). Absorptance was calculated as $A=1-R-T$.

Angle-resolved reflectance was measured by spectrally resolving the objective back focal plane using the $60\times$, NA = 0.7 objective. The position along the back focal plane corresponds to the polar incidence angle $\theta$, allowing the wavelength- and angle-dependent reflectance to be recorded simultaneously. Angle-resolved transmission was measured using a $20\times$, NA = 0.4 objective to improve angular sampling, providing an accessible angular range of approximately $-12^{\circ}$ to $+12^{\circ}$. Forward and backward measurements were acquired by flipping the sample without changing the illumination settings.

Optical micrographs of the encoded patterns were acquired using a $10\times$ objective (NA = 0.25) in reflection and transmission under broadband illumination from a 100 W halogen lamp system (Nikon TI-PS100W/A). Images were recorded using a Prosilica GC 1600 CCD camera (Allied Vision). Spectrally filtered reflection images were acquired using a long-pass filter (Carl Zeiss, 447745-8001) with a cut-on wavelength of 590 nm. The QR-code-like pattern in Figures~\ref{fig:patterns}b,c was imaged with an exposure time of $50~\mathrm{ms}$ and a detector gain of 1. The functional meta-QR code in Figure~\ref{fig:patterns}e was imaged with an exposure time of $75~\mathrm{ms}$, a detector gain of 1, and a nominal lamp-power setting of $40\%$. QR-code readability was tested by scanning the acquired optical images using ten smartphones, including both iPhone and Android devices. All ten devices successfully decoded the forward-reflection image, whereas none decoded the backward-reflection image acquired under identical conditions. Within each forward--backward comparison, the sample was physically flipped while the illumination level, objective, exposure time, detector gain, and remaining acquisition settings were kept unchanged. Additional exposure controls and transmission measurements were performed using the same meta-QR code, as described in Supporting Figure~13.

Cross-sectional transmission electron microscopy specimens were prepared using a dual-beam focused ion beam/scanning electron microscope (FIB/SEM; Zeiss NVision 40). Before milling, the region of interest was protected by an approximately $1~\mu\mathrm{m}$-thick carbon layer deposited by ion-beam-assisted deposition. Lamellae were milled using a $30~\mathrm{kV}$ Ga$^+$ ion beam, lifted out, and transferred onto Cu TEM grids. The resulting cross sections were examined by high-angle annular dark-field scanning transmission electron microscopy.

\subsection*{Data Analysis and Figure Preparation}

Experimental data were processed and analyzed using custom MATLAB and Python scripts. Image analysis was performed using ImageJ and figures were assembled using Adobe Illustrator. OpenAI ChatGPT (GPT-5.6 Thinking) was used for language editing and coding assistance. The Table of Contents graphic was prepared using the image-generation tool integrated into ChatGPT, based on sketches and scientific direction provided by the authors. All AI-assisted outputs were reviewed and verified by the authors and AI tools were not used to generate or interpret experimental data.


\newpage
 \section*{Data Availability Statement} \label{sec:data}
Data supporting the findings of this study will be available at Zenodo.
\section*{Supporting Information} \label{sec:supportinginfo} 

Nanofabrication process; ellipsometrically measured optical constants; control simulations without ITO; planar thin-film controls; sidewall-taper analysis; structural parameter sweeps and quantification of directional absorptance contrast; locally optimized geometries for maximum directional contrast and near-unity absorption; simulations using alternative materials; polarization- and azimuth-resolved reflectance simulations; angle-dependent transmission; angular response of a control metasurface without a residual a-Si film; and transmission readout, exposure controls, Michelson contrast analysis, and grayscale images of the meta-QR code.
\section*{Acknowledgements} \label{sec:acknowledgements}
 E.N.D., O.C.K, D.D.  and  G.T. acknowledge the SNSF Eccellenza  Grant PCEGP2-194181.  The authors also acknowledge the support of the following experimental facilities at EPFL: Centre for MicroNanofabrication (CMi) and EPFL: Interdisciplinary Centre for Electron Microscopy (CIME). The authors would like to thank Dr. Lucie Navratilova for TEM lamella preparation with FIB-SEM, Dr. David Reyes for the acquisition of the TEM images and data interpretation. Finally, we thank Miss Gloria Davidova for assistance with optical microscopy images and Dr. Agostino di Francescantonio and Mr. Milad Sabzehparvar for discussion on the setup. 
\section*{Author contributions} \label{sec:contributions}
G.T. and E.N.D. conceptualized the project. E.N.D. performed numerical simulations, fabricated devices, characterized samples, performed optical measurements and data analysis. D.D. assisted nanofabrication. O.C.K. assisted numerical simulations and optical characterization. E.N.D. wrote the manuscript with input from all authors. G.T. supervised all aspects of the project.
\newpage
\bibliographystyle{unsrtnat}
\bibliography{references}


\newpage
\section*{Table of Contents Graphic} \label{sec:toc}

\begin{figure}[H]
    \centering  
\includegraphics[width=12cm]{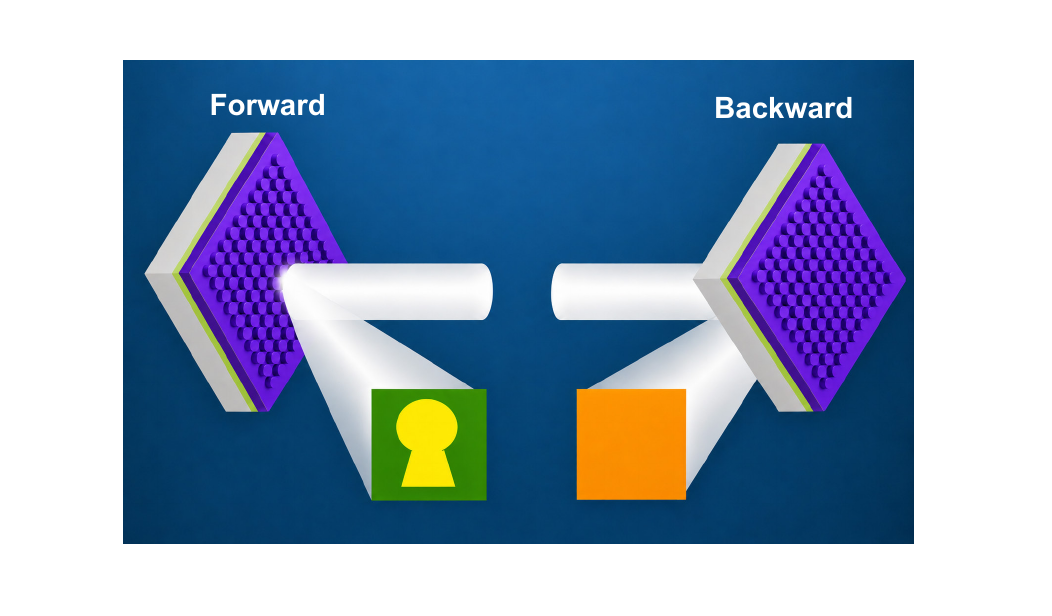}
\label{fig_toc}
\end{figure}

\end{document}